\documentclass[11pt,titlepage]{article}
\usepackage{graphicx}
\usepackage{amssymb}
\usepackage{amssymb, amsmath}

\newcommand{\at}{\makeatletter\smallish{@}\makeatother}
\newcommand{\minus}{{\scriptstyle -}}
\newcommand{\plus}{{\scriptstyle +}}
\newcommand{\thalf}{\tfrac12}

\font\smallish=cmr8 scaled \magstep1
\font\grande=cmr10 scaled \magstep4
\font\medio=cmr10 scaled \magstep2
\outer\def\beginsection#1\par{\medbreak\bigskip
      \message{#1}\leftline{\bf#1}\nobreak\medskip
\vskip-\parskip
      \noindent}

\def\laq{\raise 0.4ex\hbox{$<$}\kern -0.8em\lower 0.62
ex\hbox{$\sim$}}
\def\gaq{\raise 0.4ex\hbox{$>$}\kern -0.7em\lower 0.62
ex\hbox{$\sim$}}
\def\beq{\begin{equation}}
\def\eeq{\end{equation}}
\def\bea{\begin{eqnarray}}
\def\eea{\end{eqnarray}}
\def\bean{\begin{eqnarray*}}
\def\eean{\end{eqnarray*}}

\def \ra {\rightarrow}

\begin{document}
\bibliographystyle {amsalpha}
\titlepage
\begin{flushright}
\vspace{5mm}
CERN-TH-2017-002 \\
TAUP-3013/17 \\

\end{flushright}
\vspace{5mm}
\begin{center}

{ \grande{A  model for pion-pion  scattering in large-N QCD }} \\

\vspace{10mm}

 \large{ Gabriele  Veneziano}
\vspace{4mm}

{\sl Theoretical Physics Department, CERN, CH-1211 Geneva 23, Switzerland} 

{\sl and} 

{\sl Coll\`ege de France, 11 place M. Berthelot, 75005 Paris, France} 

\vspace{6mm}

 \large{ Shimon Yankielowicz}
\vspace{4mm}

{\sl Raymond and Beverly Sackler School of Physics
Tel-Aviv University, Ramat-Aviv 69978, Israel
 } 
 
 \vspace{6mm}
 
 \large{ Enrico Onofri}
\vspace{4mm}

{\sl Department of Mathematical, Physical and Computer Sciences} \\
{\sl Universit\`a di Parma, Parma, Italy}\\
{\sl and} \\
{\sl I.N.F.N., Gruppo Collegato di Parma, Parma, I-43124}

\end{center}

\vskip 5mm 
\centerline{\medio Abstract} 
Following up on recent work by Caron-Huot et al. we consider a generalization of
the old Lovelace-Shapiro model as a toy model for $\pi \pi$ scattering
satisfying (most of) the properties expected to hold in ('t Hooft's) large-$N$
limit of massless QCD. In particular, the model has asymptotically linear and
parallel Regge trajectories at positive $t$, a positive leading Regge intercept
$\alpha_0 < 1$, and an effective bending of the trajectories in the negative-$t$
region producing a fixed branch point at $J=0$ for $t<t_0<0$. Fixed (physical)
angle scattering can be tuned to match the power-like behavior (including
logarithmic corrections) predicted by perturbative QCD:
$A(s,t) \sim s^{-\beta} \log(s)^{-\gamma} F(\theta)$.  Tree-level unitarity
(i.e. positivity of residues for all values of $s$ and $J$) imposes strong
constraints on the allowed region in the $\alpha_0\mbox{-}\beta\mbox{-}\gamma$
parameter space, which nicely includes a physically interesting region around
$\alpha_0 = 0.5$, $\beta = 2$ and $\gamma = 3$.  The full consistency of the
model would require an extension to multi-pion processes, a program we do not
undertake in this paper.
\newpage

\section{Introduction}

In a recent paper Caron-Huot et al. \cite{Caron-Huot:2016icg} have discussed
some universal features that $2 \to 2$ mesonic scattering amplitudes should obey
under some reasonable assumptions on the properties of large-$N$ QCD.  One
particularly amazing consequence of their analysis is that the scattering
amplitude $A(s,t)$ has a universal behavior in the unphysical region
$s, t \to + \infty$ with their ratio held fixed. In this limit $A(s,t)$ should
blow up exponentially $A(s,t) \sim \exp(s\, f(t/s))$ where, furthermore, the
function $f(t/s)$ coincides (modulo a proportionality constant) with the
function appearing in string theory's 4-point function
\cite{Veneziano:1968yb}\footnote{Amusingly, the same function, rescaled by a
  factor $(1+h)^{-1}$, appears at any string-loop order \cite{Gross:1987kza}
  where $h$ is the number of loops.}

\beq \label{ftheta} 
s\, f\left(\frac{t}{s}\right) = \alpha' \left(-s
  \log s - t \log t + (s+t) \log (s+t)\right) = \alpha' \left(s \log
  \frac{s+t}{s} + t \log \frac{s+t}{t} \right)\; .  
\eeq 

In \cite{Caron-Huot:2016icg} it is also shown that the Regge trajectories
$\alpha_i(t)$ are linear and parallel at large positive $t$ in agreement, again,
with string-theoretic amplitudes.

As pointed out in \cite{Caron-Huot:2016icg} similar results cannot possibly hold
in the physical Regge and fixed-angle regions in the $s$-channel
($s \to + \infty$ with $t <0$ and either fixed or of $O(-s)$). While string
amplitudes are exponentially suppressed at fixed angle (or down by arbitrarily
high powers of $s$ at sufficiently negative $t$), we expect an inverse power-law
behavior (up to logs) in $QCD$ because of asymptotic freedom.  In other words
some kind of Stokes phenomenon must intervene and drastically change the
asymptotic form (\ref{ftheta}) as one moves in the asymptotic $s, t$ complex
planes.

The challenge, therefore, is to be able to achieve such a result while keeping
all the properties expected to hold at large-$N$, in particular, the pole-like
nature of all the singularities. An old argument due to Mandelstam strongly
suggests that Regge trajectories should be analytic functions of their argument
and that the absence of branch point singularities in the Mandelstam variables
should imply that $\alpha_i(t)$ are actually entire functions of $t$, hence,
essentially, that they must be linear. Although Mandelstam's is not a rigorous
argument and some ways out of its conclusion have been discussed in
\cite{Caron-Huot:2016icg}, constructing Regge-behaved amplitudes with complex
Regge trajectories and no branch points in the Mandelstam variables appears to
be extremely hard.

Accepting instead Mandelstam's argument it looks that the only possibility one
can play with is to give up the universality of the Regge slopes. If
trajectories of lower slope are added they will not affect much the positive-$t$
behavior while they will drastically modify the one at negative $t$ (Regge and
fixed angle alike).  We note here that the idea of adding to the usual
string-theoretic amplitudes with universal slope $\alpha'$ other amplitudes with
fractional slopes $\alpha'/k$ had been already proposed by Andreev and Siegel
\cite{Andreev:2004sy}. Their motivation was somewhat similar to ours. However,
those authors did not consider in any detail the important constraint of
(tree-level) unitarity.

In this note we would like to present a class of toy models for $\pi \pi$
scattering incorporating this idea and impose on them the whole set of
constraints we believe to apply to this process in the large-$N$ limit.  The
outline of the paper is as follows: In Sect.~2 we review the large-$N$-QCD
expectations for $\pi \pi$ scattering.  In Sect.~3 we present a generalization
of the old Lovelace-Shapiro (LS) model \cite{Lovelace:1969se, Shapiro:1969km}
and then consider its various properties, in particular, its asymptotic limits
and the non-trivial constraints from tree-level unitarity.  This latter
constraint is analyzed in detail (mainly numerically) in Sect.~4 where we
identify a region in a $3$-dimensional parameter space where unitarity is
satisfied.  In Sect.~5 we summarize our results and discuss the limitations of
our approach essentially due to our inability to generalize the method to
higher-point functions or, alternatively, to processes involving the massive
intermediate states as external legs. Details about the numerical
approach of Sect.~4 are given in the Appendix. 
  
 \section{Large-$N$-QCD expectations for $\pi \pi$ scattering}
 
 We will be considering the large-$N$ limit of the QCD scattering amplitude
 $A_{\pi \pi \ra \pi \pi}$ in the chiral (i.e massless quark) limit. We recall
 that large-$N$ QCD preserves (actually reinforces) asymptotic freedom.  We
 assume, as usual, that large-$N$ QCD is a confining theory and that its chiral
 symmetry is realized \`a la Nambu-Goldstone (NG), with the pions representing
 the corresponding massless NG bosons. As a consequence, isospin symmetry is
 exact and each $\pi \pi \ra \pi \pi$ scattering amplitude can be decomposed in
 pure (s-channel) isospin amplitudes $A_I(s,t), I = 0,1,2$. Furthermore, all
 amplitudes must exhibit an ``Adler zero" i.e. they must vanish when any one of
 the external pion momenta goes to zero.  In what follows we first discuss the
 further properties that a viable model for such a 4-point amplitude should
 satisfy and then make a few further assumptions.
 
  \subsection{Crossing symmetry}

  There is no reason to doubt that large-$N$ QCD obeys crossing
  symmetry. Hence a single analytic function will describe the three
  processes in which either $s$, $t$ or $u$ is positive and the two
  other complementary Mandelstam variables are negative (and smaller
  in absolute value than the positive one).
 
 \subsection{Meromorphicity}
 
 We know that, because of confinement and large-$N$, the scattering
 amplitude should only have pole-like singularities lying at discrete
 positive values of the three Mandelstam variables $s, t, u$ (with
 $s+t+u =0$).  Interactions are down by powers of $1/N$ in the
 large-$N$ limit, the 4-point function being of order $1/N$.
 
 Furthermore, the planar topology implied by the large-$N$ limit tells
 us that the different amplitudes must be linear superpositions (with
 known coefficients) of amplitudes that can be obtained from a single
 one, say $A(s,t) = A(t,s)$, by permutations of the Mandelstam
 variables. In particular:
 \beq
 A_0 = \tfrac{3}{2}[A(s,t) + A(s,u)] -
 \thalf A(t,u) ~~;~~ A_1 = A(s,t) - A(s,u)~~;~~ A_2 = A(t,u)\; , 
\eeq
 where each amplitude has poles in the Mandelstam variables explicitly
 shown as arguments but not in the third one.  The Adler zero implies
 $A(0,0) =0$.
 
  \subsection{Tree-level unitarity}
  
  QCD is a unitary quantum field theory and, as such, we expect it's large-$N$
  limit to obey certain unitary constraints. The 't Hooft large-$N$ limit
  \cite{tHooft:1973alw}, as opposed to the one proposed by one of us
  \cite{Veneziano:1976wm}, corresponds to a tree-level approximation in terms of
  color-singlet bound states. At that level unitarity demands that the residue
  of each $s$-channel pole --and at each value of angular momentum $J$ and
  isospin $I$-- should be positive.
  
     \subsection{Asymptotic freedom and the fixed-angle, high-energy limit}
   
     Since large-$N$ QCD is asymptotically free this should reflect itself in
     the fixed angle (large s, large t with $s/t$ fixed) scattering. Note that
     the physical region corresponds to negative $t$. The non-physical region of
     positive t (imaginary fixed-angle scattering) is the one in which a
     universal behavior was established in \cite{Caron-Huot:2016icg}.  Of course
     this universal behavior should be part of any viable model for the 4-point
     amplitude.

     The counting rules of Brodsky-Farrar \cite{Brodsky:1973kr} (see also
     \cite{Lepage:1980fj}) imply\footnote{We are grateful to Mark Strikman for
       an interesting exchange on the present status of exclusive processes in
       QCD.}:
     \beq
     \label{fixedaQCD}
  A(s,t) \ra s^{2 -n_q/2} F(\theta)\; ,
  \eeq
  as $s \ra + \infty$, $t \ra - \infty$ and $\cos\theta = 1 + \frac{2t}{s}$
  fixed. Here $n_q$ is the total number of valence quarks participating in the
  process (hence, at least naively, $n_q =8$ in the process at hand). Because of
  the logarithmic running of the strong coupling in QCD some inverse powers of
  $\log s$ are also expected.
  
  A more subtle question is whether the above prediction holds for the large-$N$
  (planar) limit of QCD. It could be that the fixed angle high energy limit does
  not commute with the large $N$ limit. In this case we should perform the
  latter limit first and the fixed angle behavior may turn out to be more damped
  than in (\ref{fixedaQCD}).

  \subsection{Regge limit and DHS duality}
  
  It is not clear whether the large-$N$ limit of QCD obeys a Regge behavior (as
  $s \ra \infty$ and $t$ is kept fixed) of the type:
  \beq
  \label{RPB}
  A(s,t) \ra \sum_i \beta_i(t) s^{\alpha_i(t)} \,,
  \eeq
  i.e. is controlled entirely by Regge poles. As we shall see our own model will
  fail to satisfy (\ref{RPB}) at sufficiently negative $t$.
 
  Our somewhat milder assumption here will be that in the Regge limit $A(s,t)$
  goes to zero if $t$ is taken to be sufficiently negative. Unfortunately, one
  cannot use directly AF to study this regime. However, by continuity with the
  fixed angle regime we have just discussed the amplitude should go to zero at
  sufficiently negative $t$.
   
  An immediate consequence of such an assumption is that, at sufficiently
  negative $t$, one can write down an unsubtracted dispersion relation for
  $A(s,t)$ with an imaginary part given by a sum of Dirac $\delta$-functions.
  Therefore, as pointed out long ago (see for instance \cite{Veneziano:1974dr}),
  the sum over the $s$-channel poles should necessarily give, by analytic
  continuation, the $t$-channel poles, which is nothing but a simple restating
  of the old Dolen-Horn-Schmid duality \cite{Dolen:1967jr}. In other words,
  under a reasonable assumption on the Regge limit of large-$N$ QCD, DHS duality
  is automatic.

 \section{A generalized Lovelace-Shapiro model}
 
 The original Lovelace-Shapiro (LS) model takes the following expression for $A(s,t)$:
 
 \beq
 \label{LS}
 \frac{A(s,t)}{g^2} =  \frac{\Gamma (1\minus\, \alpha(s))\,\Gamma (1\minus\, \alpha(t))}{\Gamma (1\minus\, \alpha(s)\minus\,\alpha(t) )} =
 (1 -  \alpha(s) - \alpha(t))\, B(1\minus\, \alpha(s) , 1\minus\, \alpha(t)) \;,\quad \alpha(t) = \tfrac12 + \alpha' t
 \eeq
 where $g$ is the string coupling constant and $\alpha'$ the (universal)
 Regge-slope parameter (later reinterpreted as the inverse of the string
 tension). Dimensional transmutation in large-$N$-QCD is able to generate a
 scale, such as $\alpha'$ while $g$ will turn out to be of order $1/\sqrt{N}$.

  \subsection{Implementing Large-$N$-QCD expectations}

  As already mentioned, according to \cite{Caron-Huot:2016icg} the large-$N$ QCD
  amplitude should be meromorphic and should exhibit, at large positive $t$,
  asymptotically linear and parallel Regge trajectories. Accepting the argument
  by Mandelstam we mentioned in the introduction (see however in
  \cite{Caron-Huot:2016icg} a discussion of possible ways out), having no cuts
  in the amplitude implies that it should be constructed from a sum of terms
  each one having just straight Regge trajectories but with different slopes.
  Clearly, any finite number of such terms will lead to an exponentially falling
  amplitude at fixed angle. For the same reason, an infinite sum should contain
  terms with arbitrarily low slopes. Finally, as we shall argue below, unless
  the possible slopes (as well as the Regge intercepts) are rationally related
  to each other, unitarity (i.e. positivity of residues) will be violated. In
  the end, we shall take a discrete sum with slopes decreasing as $1/k$ (i.e. of
  tensions growing like $k$) precisely as in \cite{Andreev:2004sy}.
  
  \subsection{A simple ansatz}
  
The previous arguments suggest taking a generic ansatz of the form:

 \bea
 \label{GLS}
 \frac{A(s,t)}{g^2} &= &\sum_{k=1}^{\infty}\, c_k\, A_k(s,t) \nonumber \\
 A_k(s,t) &=& -\frac{(s+t)}{k}\, B\left(1\minus\, \frac{1-a+s}{k} ,
   1-\frac{1-a+t}{k}\right)\,,\; 0 \le a \le 1 \; .  \eea We see that
 the term with $k=1$ reproduces the LS amplitude (\ref{LS}) if we set
 $a=\frac12$ and use units of energy such that $\alpha' = 1$. On the
 other hand our ansatz Eq.\eqref{GLS} implements the Adler zero for any
 value of $a$.
 
 The pole structure of Eq.\eqref{GLS} in $s$ is simple. Poles of $A_k$ are
 located at $s = a + (k\minus 1) + k q, (q = 0, 1, \dots) $. Therefore the poles
 of $A$ are only those present in $A_1$, i.e. they lie at $s = a + M$ with
 $M = (k\minus 1) + k q, (M = 0, 1, \dots) $. Clearly $A_k$ can contribute to
 the residue at the $M$th pole only if $q$ is a divisor of $M\plus1$. Thus also,
 necessarily, $k \le M\plus1$ so that only a finite number of terms in the sum of
 Eq.\eqref{GLS} contributes at a given mass level. If $k$ is a divisor of
 $(M\plus 1)$ we have: 
\beq
 \label{Rest}
 {\rm Res}A_k(s=a+M) = \frac{a +M +t}{q_k!} \,\frac{\Gamma(q_k +
   (1\minus a+t)/k)}{\Gamma((1\minus a+t)/k)} \,,\; q_k
 \equiv \frac{M\plus 1}{k} \minus  1\; .
 \eeq

 The above residue is a polynomial of degree $(q_k+1)$ in $t$ and can therefore
 be rewritten as polynomial of the same degree in
 $x = \cos \theta = 1 + \frac{2t}{a + M}$. When the result of this substitution
 is inserted in our ansatz (\ref{GLS}) we find:
 \begin{equation} \label{ResAx} 
   g^{\minus 2}{\rm Res}A\vert_{s=a+M} = \frac{ (a
     +M)(1+x)}{2}\times { \sum_k}^\prime\,\frac{ c_k}{q_k!} \left( (1+\thalf(
     a\,(x\minus 3) + M\,(x\minus 1))/k\right)_{q_k}
 \end{equation}
 where we recall that the sum over $k$ runs over the divisors of
 $(M\plus 1)$ and $(a)_n \equiv\Gamma(a+n)/\Gamma(a)$ denotes the
 Pochammer's symbol.

\subsection{Asymptotic limits}

We will now discuss the high-energy limit of our amplitude Eq.\eqref{GLS}.  The
novelty, with respect to standard string amplitudes, will be a kind of Stoke's
phenomenon in the sense that the $s \ra + \infty$ behavior changes drastically
as one moves from the physical $t <0$ region to the unphysical one $t >0$. This
will be the case for both the Regge (fixed $t$) regime and for the fixed angle
(fixed $t/s$) one. We shall now discuss the two regimes in turn.

\subsubsection{The fixed-angle limit}\label{sec:fixed-angle-limit}

Obviously the fixed-angle behavior of $A_k$ parallels the one of
$A_1$, Eq.\eqref{ftheta}: 
\beq
\label{Fixeda}
A_k(s,t) \ra \exp\left( \minus\, \frac{ s \log s + t \log t -(s+t) \log (s+t)}{k}
\right)\equiv \exp (\minus B/k) \, .  \eeq In order to
estimate the asymptotic behavior of the sum (\ref{GLS}) we convert it into an
integral over a continuous variable $\lambda = 1/k$: 
\beq\label{Fixedaint}
A(s,t) \sim \int_0^1 \frac{d \lambda}{\lambda^2}\, c(\lambda^{-1})
\exp(- \lambda B) = B \int_0^B \frac{d x}{x^2}\, c(B/x) \exp(-x)\, .
\eeq
In the unphysical region ($s,t >0$) $B$ is negative and the integral is dominated
by $x \sim B$ so that, essentially $A \sim A_1$. In the physical fixed
angle region $B >0$ and the integral, dominated by the small-$x$
region gives: 
\beq
\label{Fixedasum}
A(s,t) \sim B \int_0^{\infty} \frac{d x}{x^2} \,c(B/x)\, \exp(-x) \sim
B^{- \beta}\, \Gamma(\beta ) \sim s^{-\beta}\; , \eeq if
$c(k) \sim k^{-\beta-1}, \beta >0$ for $k \ra \infty$.  More
generically: \beq A(s,t) \sim B \,c(B) \eeq so that it can include,
for instance, logarithmic corrections to an inverse-power behavior.
Note that our model predicts a precise angular dependence.  Taking at
face value the quark-counting prediction, would require
$c(k) \ra k^{-3} (\log k)^{-\gamma}$, presumably with $\gamma \sim 3$
\footnote{The latter estimate comes from observing that the minimal
  number of planar gluons needed to induce the hard scattering at the
  partonic level generates three powers of $\alpha_s$ at the 
  hard scale $s$. It ignores, however, possible extra corrections coming from
  the pion wave functions.}.
 
\subsubsection{The Regge limit}
The individual $A_k$ amplitude has the following Regge limit: 
\beq
\label{Regge}
A_k(s,t) \ra \Gamma\left(1-\frac{1-a+t}{k}\right)\,\left(
  -\frac{s}{k}\right)^{(1\minus a+t)/k} \, .
 \eeq 
 Thus $(1\minus a)/k$ is the Regge intercept of the $k$th amplitude.
 
In order to estimate the Regge behavior of the sum (\ref{GLS}) we proceed as before:
\begin{align}
  \label{Reggeint}
A(s,t) &\sim \int_0^1 \frac{d \lambda}{\lambda^2}\, c(\lambda^{-1})\,
\exp\left(\lambda(1-a+t) \log (\lambda s)\right) \\\nonumber
&= s \int_0^s \frac{d x}{x^2}\, c\left(s/x\right) \exp\left(x
  \frac{(1-a+t)}{s} \log (x)\right)\, .
\end{align}
In this case we have to distinguish the region $t > a-1$ (which includes a small
physical region around $t=0$ as well as the whole unphysical region $t>0$) from
the ``large'' negative-$t$ region $t < a-1$.

In the former case, the exponential in Eq.\eqref{Reggeint} is large and, once
more, $A \sim A_1$ with the typical stringy Regge behavior. In the latter case,
however, the small-$x$ region dominates.

Taking again, as a typical example, the case $c(k) \sim k^{-\beta-1}, \beta >0$
for $k \ra \infty$, we find:
\beq
\label{Reggesum}
A(s,t) \sim \left[ (a\minus 1\minus t) \,\log \left(\frac{s}{a\minus 1\minus t} \right)\right]^{- \beta}\, , 
\eeq 
a result matching nicely the fixed angle behavior of
Eq.\eqref{Fixedasum} 
when $-t$ becomes $O(s)$.
 
From the point of view of Regge theory this behavior corresponds to a cut in
angular momentum situated at $J=0$ and of the form: 
\beq
 \label{Jsing}
 A(J) \sim J^{\beta-1} \log(J) ~~;~~ A(J) \sim J^{\beta-1}\, ,
  \eeq 
  for integer and non-integer $\beta$, respectively. We thus conclude that the
  sum over an infinite number of Regge poles has produced a branch point
  singularity (``Regge cut") at $t = a\minus 1$ and $J=0$.

\begin{figure}[ht]
   \centering
   \includegraphics[width=1.\linewidth]{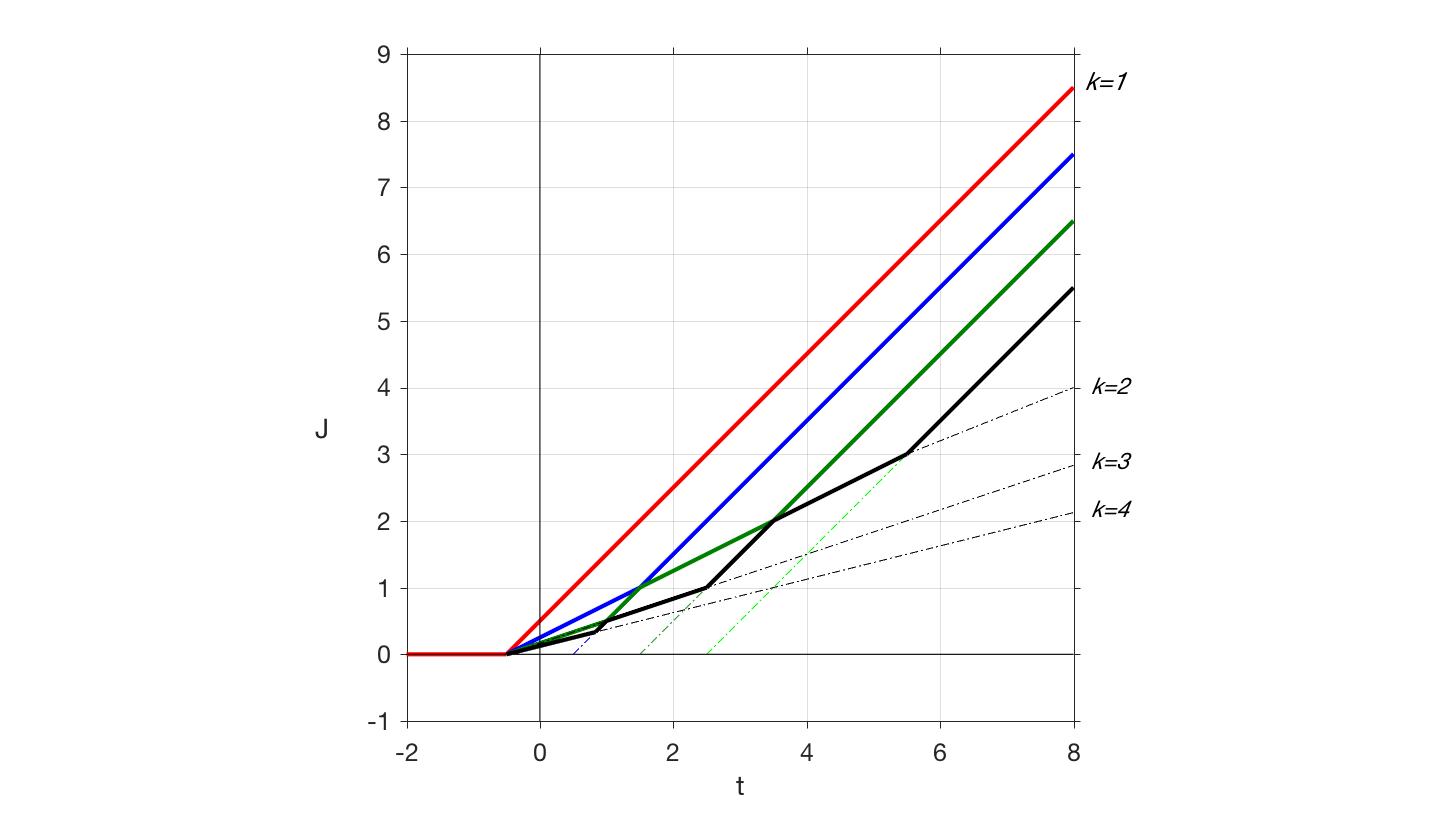}
   \caption{{\smallish Leading (in red) and next three subleading (in
       blue, green, black) Regge trajectories of our model for $\alpha_0
       = \thalf$. The sum of Regge poles turns into a branch point at
       $J=0$ below $t = -\thalf$.}} 
   \label{fig:Regge}
 \end{figure}

  The Regge pole structure at $t > a\minus 1$ (shown in Fig.~\ref{fig:Regge}) is
  also noteworthy.  The leading trajectory is the one given by $A_1$. The
  next-to-leading trajectory is again the one of $A_1$ for $t >a+1$, but,
  between $a-1$ and $a+1$ it is given by the leading trajectory of $A_2$. There
  is a break in the slope and, exactly at $t = a+1$, there is a double pole at
  $J=1$.

 The structure becomes more and more complex as we go to even lower Regge
 poles. They always start asymptotically linear and parallel (as argued in
 \cite{Caron-Huot:2016icg}) with slope $\alpha'$, but then they go through a
 series of breaks simulating a curved trajectory while remaining piecewise
 linear. This may turn out to be the only way to implement meromorphicity. Of
 course, we expect that, once higher orders in the large-$N$ expansion are
 included as in \cite{Veneziano:1976wm}, the resulting trajectories will be
 smooth and curved and, correspondingly, resonances will acquire a finite width.

\subsection{Tree-level unitarity}

The most stringent constraint comes at this point from requiring tree-level
unitarity which becomes nothing but positivity of the residues of each pole in
$s$ for each definite $s$-channel angular momentum and isospin (these being the
quantum numbers diagonalizing the unitarity constraint).

Note that the $I=2$ amplitude has no $s$-channel poles while the $I=0$ and $I=1$
amplitudes have only even and odd angular momenta. It thus turn out that a
necessary and sufficient condition for tree-level unitarity is the one that can
be imposed on $A(s,t)$ itself.

In the next section we will present the results we have been able to obtain so
far on the unitarity issue.

\section{Tree-level unitarity constraints}

The issue of positivity of pole residues for all masses and spin is, to the best
of our knowledge, an unsolved one even for the old LS amplitude. What is known
in that case is that positivity is lost as soon as one moves away from a
critical value $\alpha(0) = \thalf$ of the intercept or from the dimensionality
$D=4$ of space-time.  Moving in the direction of a lower intercept and/or of a
higher $D$ immediately turns a zero-norm state into a ghost.  Although we have
not found a rigorous proof, we have many numerical and analytic arguments in
favor of positivity for all masses and spins in the $D=4, \alpha(0) = \thalf$
case.

If we now move to our ansatz Eq.\eqref{GLS} for $D=4$, it is clear
from what precedes that each individual amplitude $A_k, \,k \ne 1$ will
generically lead to negative residues because of their lower and lower
intercept as $k$ is increased\footnote{For this reason negative
  residues would appear had we not chosen quantized values for the
  Regge slopes.}. It will be therefore quite non trivial to achieve
positivity for $D=4$ and $\alpha(0)$ in the neighborhood of $\thalf$.

\subsection{Analytic results}\label{sec:analytic-results}
  
  Being able to prove analytically that, in a suitable region of
  parameter space, the positivity requirement is satisfied looks
  highly improbable. On the other hand one can certainly check
  positivity for the very first few levels and numerically for many of
  them (see below). The hope then is to be able to prove that if
  positivity holds up to a large mass level (and nay spin) it will
  keep holding all the way to infinitely large masses and spins.
  After all, while the first levels are deeply in the quantum regime,
  the high-mass and high-spin level should have a (semi)classical
  counterpart. And we do not expect unphysical properties in a
  classical string.
  
  We have then looked at this regime by smoothing out the imaginary
  part of the amplitude (which gives nothing but its Regge limit) 
  \beq
  \label{ImRegge}
  {\rm Im} A_1 = \pi s^{\alpha_0} \left[ \frac{s^{t}
    }{\Gamma(\alpha_0 + t)} \right] \; ,
  \eeq 
  and by replacing the
  Legendre projection by the well known passage from momentum
  transfer to impact parameter $b$ to be identified at large $s$ and
  $J$ with a continuous value\footnote{The fact that both the mass
    and the spin of the intermediate states become continuous
    reinforces the belief that such a procedure has to do with taking
    some classical limit.} for $2J/\sqrt{s}$.
  
  It turns out that such a procedure can be easily implemented and
  that the needed Fourier-Bessel transform can be well approximated
  by a saddle point in the complex $t$ plane giving: \beq
  \label{saddle}
  \log\left( \frac{1}{\pi} \Gamma(\alpha_0)s^{-\alpha_0} {\rm Im}
    A_1(s,b)\right) = - \frac{b^2}{4 W} + \frac{b^2 Y}{16 W^2}- \log
  \left[ \frac{\Gamma(\alpha_0 +\frac{b^2 }{16 W^2}
      )}{\Gamma(\alpha_0)}\right]\; , 
      \eeq 
      where
  $W \equiv - W_{-1}(-\frac{J}{2s})$ is the lower branch of Lambert
  (product-log) function solving the equation: \beq
  \label{Lambert}
  \log W(x) - W(x) = \log(x) 
  \, .
  \eeq
  
\begin{figure}[ht] 
\begin{center}
\includegraphics[width=0.6\linewidth]{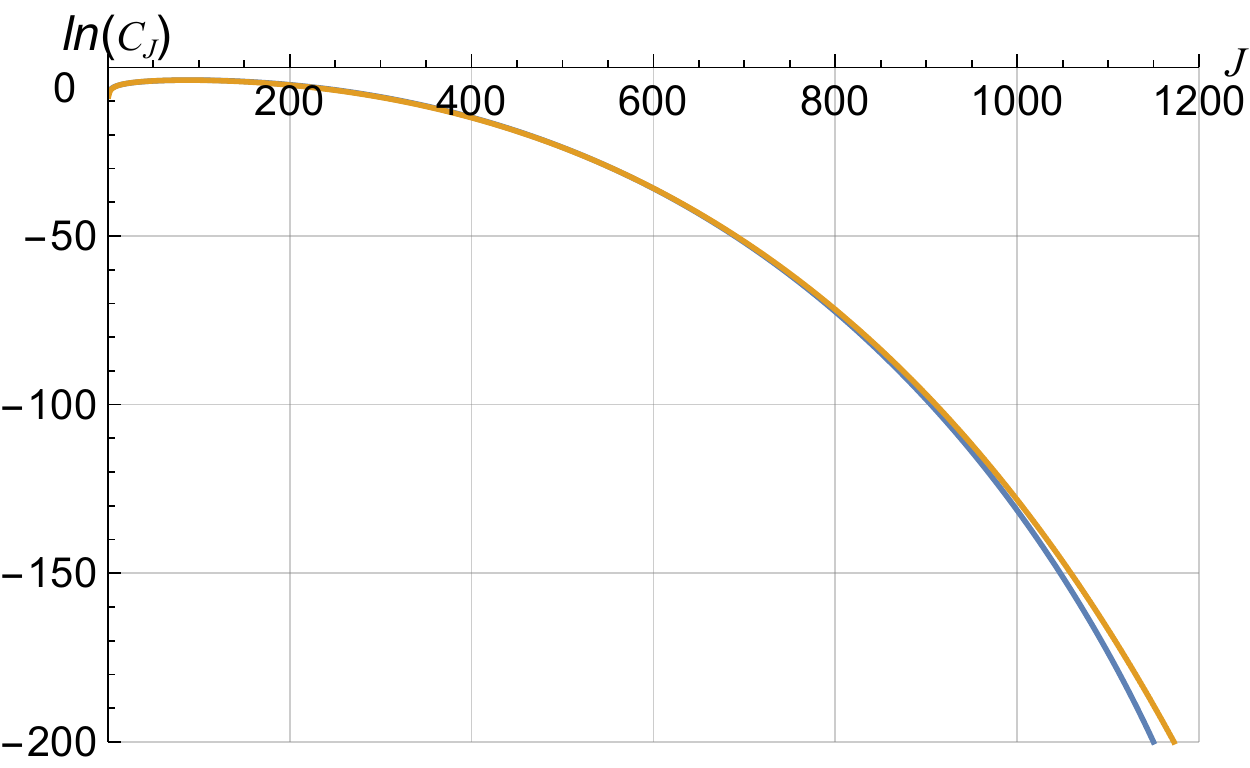}
\end{center}
\caption{{\smallish Log-plot of Legendre expansion coefficients $C_J$ by the saddle
    point (blue line) and by the algebraic method of
    Sec.~\ref{sec:algebraic-algo} up to {\footnotesize $J=1200$.}}}
\label{fig:saddle} 
\end{figure}

  We have checked this expression against the numerical ones described
  below and found an amazingly good agreement over more than $20$
  orders of magnitude (see Fig.~\ref{fig:saddle})\footnote{Curiously
    enough, an isolated leading saddle only exists for $J < (2/e)\,s$
    (while we need $J \le s$ to cover all possible values). For
    $J > (2/e)\,s$ a second saddle collides with the first and then
    the two move away for the imaginary axis. Presumably, the saddle
    point method can be suitably extended to this regime but we have
    not attempted to do it so far.}.  An advantage of this method is
  that it gives an explicitly positive result in its region of
  validity.

  One should then consider our superposition in Eq.\eqref{GLS}ñ of
  quantities like the one in (\ref{saddle}) and check what conditions
  on the $c_k$ emerge from the positivity requirement.  Using
  (\ref{Regge}) the quantity to be considered is: \beq
  \label{saddlesum}
  {\rm Im} A = \pi \sum_k c_k \left(\frac{s}{k}\right)^{ \alpha_0/k} \left[ \frac{(\frac{s}{k})^{t/k} }{\Gamma(\frac{\alpha_0 + t}{k})}  \right]\; ,
  \eeq
  and therefore the needed generalization of (\ref{saddle}) is:
  \beq
  \label{saddlek}
  \log\left( \frac{1}{\pi k} \Gamma(\frac{\alpha_0}{k})s^{-\frac{\alpha_0}{k}} {\rm Im} A_k(s,b)\right)  =
  - \frac{k b^2}{4 W_k} + \frac{b^2 Y}{16 W_k^2}-
  \log \left[ \frac{\Gamma(\frac{\alpha_0}{k} +\frac{k b^2 }{16 W_k^2} )}{\Gamma(\frac{\alpha_0}{k})}\right] \; ,
  \eeq
  where $W_k \equiv - W_{-1}(-\frac{kJ}{2s})$.

  The results obtained by this method are shown in
  Fig. \ref{fig:saddle} and can be compared with the ones computed by
  the methods of next section\footnote{This comparison refers to the
    $k=1$ amplitudes with $\alpha_0=\tfrac23$.}. We plot the expansion
  coefficients $C_J$ in Legendre polynomials (not to be confused with
  the weights $c_k$ of Eq.\eqref{ResAx}) in logarithmic scale.
  Agreement is even better with the numerical results obtained using
  Eq.\eqref{saddlesum}.

   Another analytic approach that we have tried, so far unsuccessfully, is of the induction type. If, above a certain $s = a + \bar{M}$, one could prove that positivity at $s = a + M$ implies positivity at $s = a + M +1$
one would be done. This looks like a hard but not impossible problem.

\begin{figure}[ht] 
\begin{center}
  \includegraphics[width=0.75\linewidth]{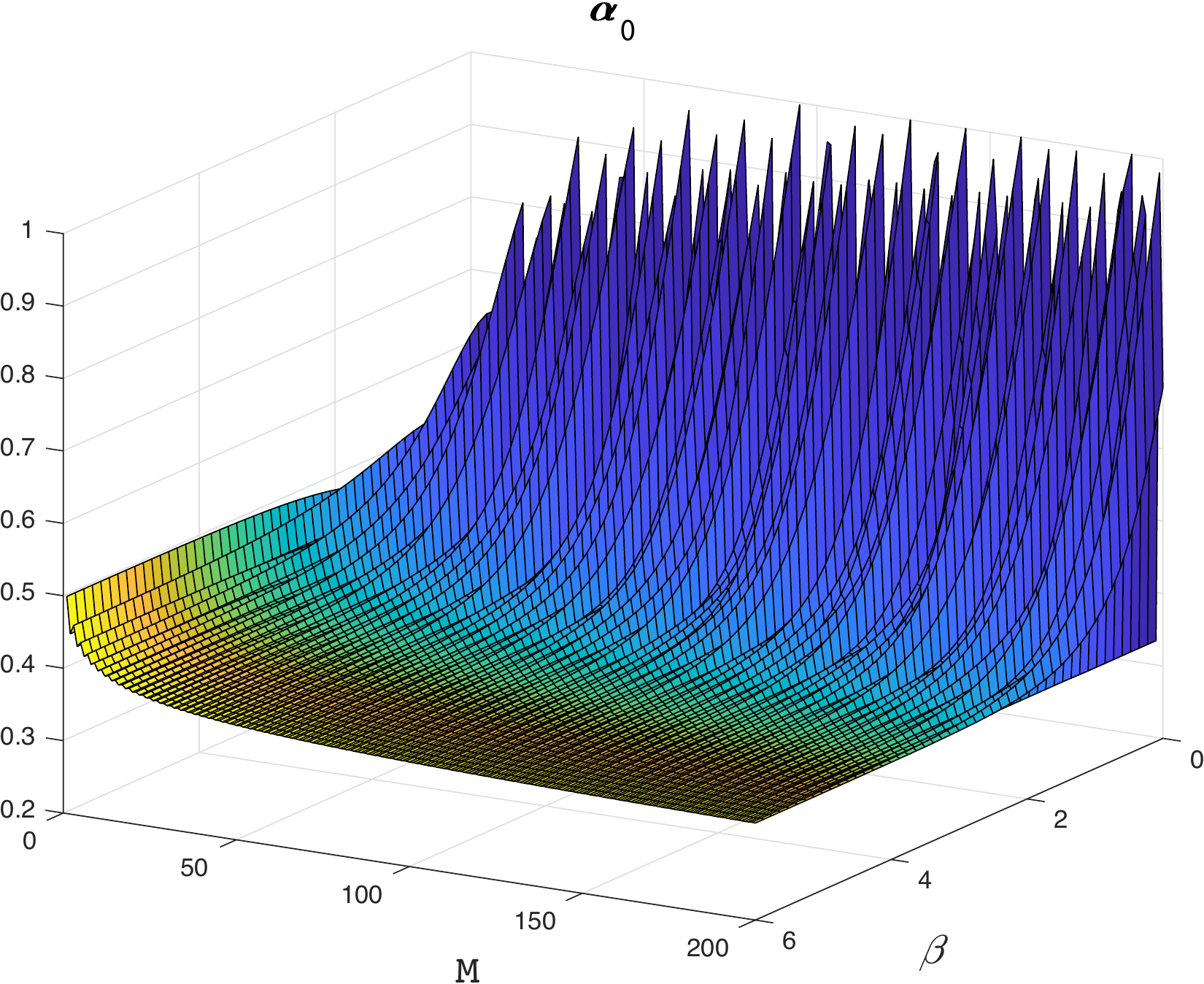}
\end{center}
\caption{{\smallish The ``critical landscape'': $\alpha_0$ values
    above the surface give positive residues (here and in the
    following {\footnotesize$\gamma=\beta+1$})}}
\label{fig:AcritAss}
\end{figure} 
\begin{figure}[h] 
\begin{center}
  \includegraphics[width=.65\linewidth]{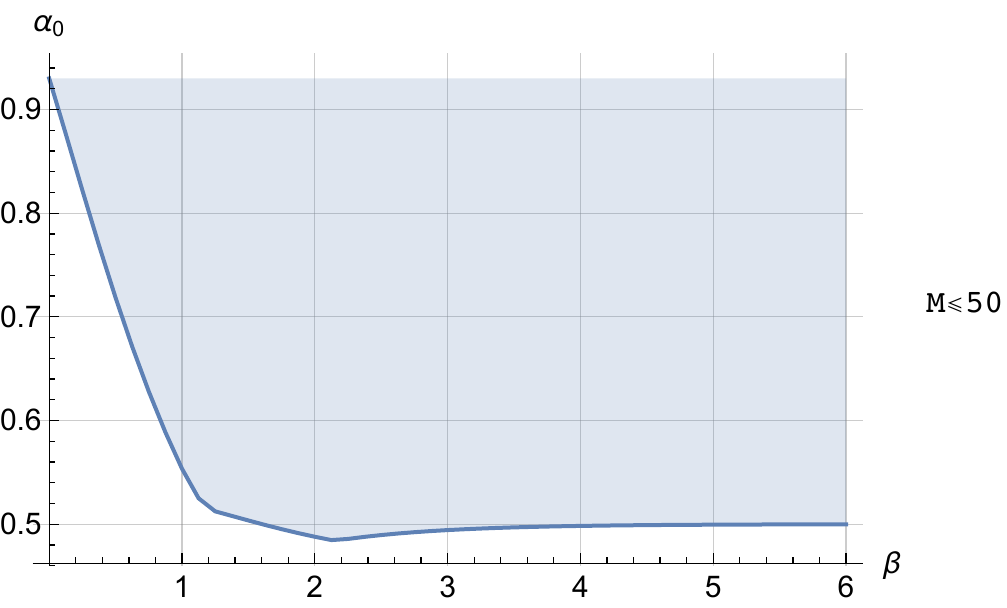}
\end{center}
\vspace{-5mm} 
\caption{{\smallish Same as the plot of Fig.\ref{fig:AcritAss}
    projected on the ${\footnotesize (\beta,\alpha_0)}$ plane. Shaded
    region corresponds to positive residues.
}}
\label{fig:shaded}
\end{figure} 

\begin{figure}[h] 
\centering
\includegraphics[width=0.75\linewidth]{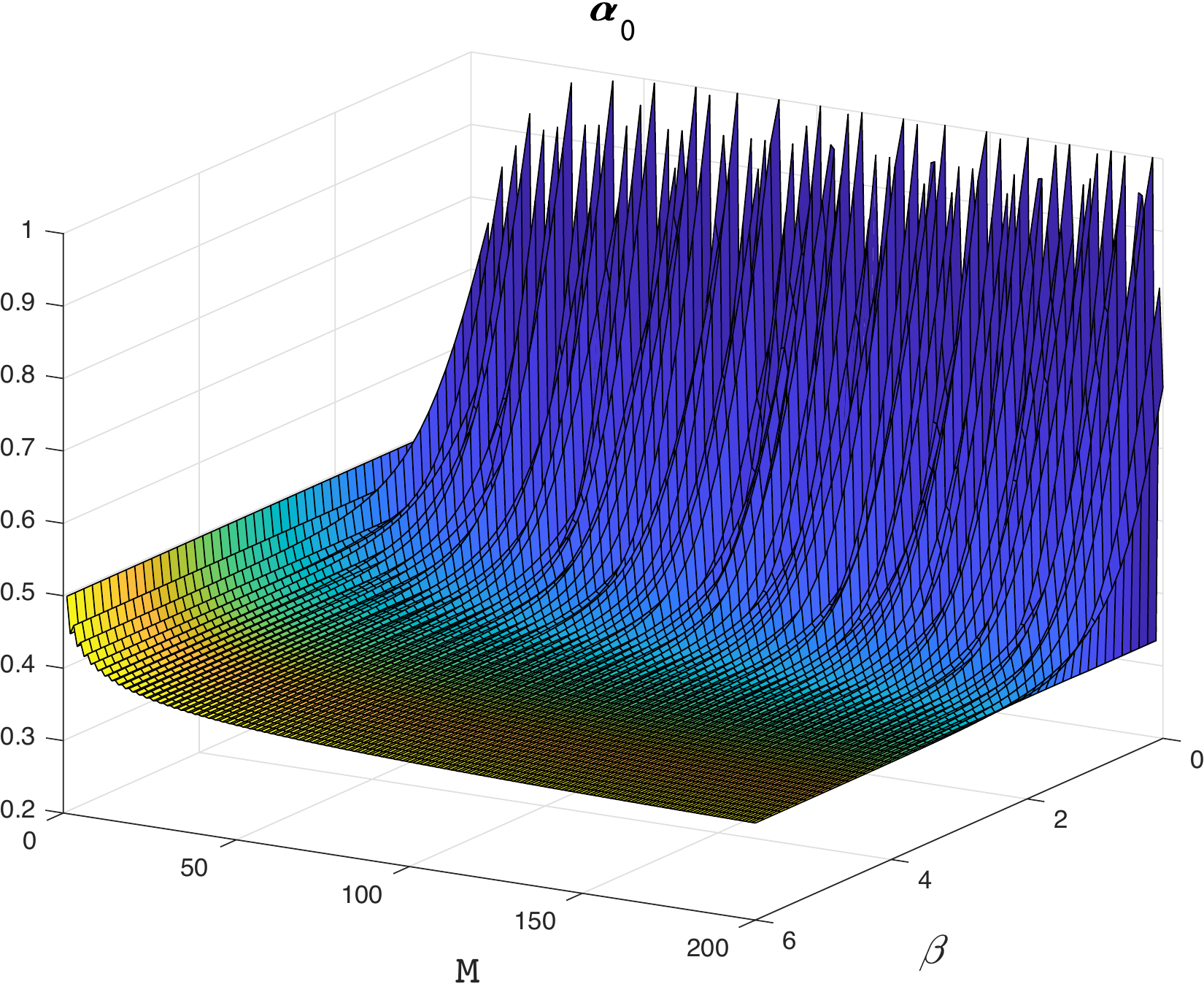}
\caption{{\smallish Same as Fig.~\ref{fig:AcritAss} when the summation
    in Eq.~{\smallish \eqref{ResAx}} is restricted to odd divisors.}}
\label{fig:PosOdd}
\end{figure} 

\begin{figure}[h] 
\centering
  \includegraphics[width=0.65\linewidth]{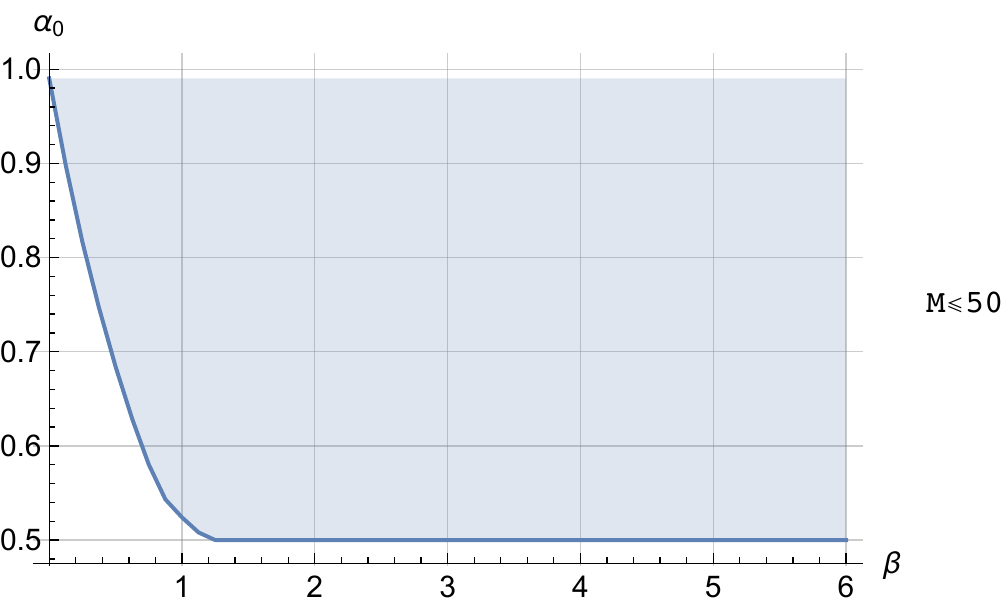}
\caption{{\smallish Same as the previous figure projected on the
    {\footnotesize $(\beta,\alpha_0)$} plane}}
\label{fig:PosOddFlat}
\end{figure} 

\subsection{Numerical results}\label{sec:numerical-results}
The expansion of the function given in Eq.~\eqref{ResAx} (with the ansatz $c_k = k^{-\beta-1}\,(1+\ln(k))^{-\gamma}$) into a series
of Legendre polynomials represents an apparently straightforward task
involving a number of integrations of the kind
$\int_{-1}^1\,f(z)\,P_n(z)\,dz$; still we have to face the problem
that {\it i\/}) the number of integrals to be computed to explore the
whole ``critical landscape'' is rather high (of the order of
$\sim 10^6$) and {\it ii\/}) high accuracy is needed , since we are
dealing with high order polynomials at the level of $O(10^3)$ as
required for the comparison with analytic data. Hence we have chosen
an alternate strategy, that is presented in the Appendix.  Here we
limit ourselves to present the results of numerical computations as
condensed in Fig.~(3-6), where it is clear that there is an ample
domain in $(\beta,\alpha_0)$ plane where the unitarity constraint is
satisfied.  In Fig.~\ref{fig:AcritAss} we present a bird's view of the
critical surface, while in Fig.\ref{fig:shaded} we report the region
in $(\beta, \alpha_0)$ plane where the expansion coefficients are
positive, which is equivalent to look at Fig.~\ref{fig:AcritAss} in
the direction of
$M$-axis. Fig.s~(\ref{fig:PosOdd},~\ref{fig:PosOddFlat}) refer to the
restriction of the sum in Eq.~\eqref{ResAx} to odd divisors of
$M\plus1$, whose motivation should be clear from the last part of the
conclusions.  Algorithmic details can be found in the Appendix.

\section{Conclusions}

Summarizing our results we should immediately remark that positivity
of residues in the massless four point function is only a necessary
condition for the unitarity of the model. Hence, not surprisingly
there is ample room for satisfying such a (still non trivial)
constraint. There is a-priori  room for hidden degeneracy underlying
poles of the 4-point amplitude and the no-ghost constraint demands
positivity for each individual contribution to the residue at each
given $s$, $J$ and $I$. To disentangle such a possible degeneracy and
check for unitarity one has to consider also higher $n$-point
amplitudes as it was done in the early days of string theory. At that
point one should also impose positivity for elastic $2\ra 2$ processes involving the massive sector of the theory, a point recently stressed by Arkani-Hamed \cite{AH}.
Having amplitudes with two external excited states would also allow to construct QCD-string loops  and check the expectation \cite{Bochicchio:2016euo} that such loops, unlike those of the Nambu-Goto string, are UV-divergent. All this, however, is far beyond the scope of this paper. 

To reconcile the universal exponential behavior of the amplitude at
large positive $s$ and $t$ and it's Regge behavior with the softer
power (up to logs) behavior associated with AF, we need, effectively,
a dominant Regge trajectory which is linear with slope 1 (in units of
$\alpha'$) at large positive $t$, but flattens out at negative
$t$. Our model amplitude effectively achieves this by having an
infinite sum of string amplitudes which differ by their slope
(associated with the string tension), with the $k$th term in the sum
having a slope of $1/k$. If we draw the leading linear trajectories
associated with all these terms we see (Fig. 1) that the effective
dominant (leading $J$ for fixed $t$) trajectory has a "kink" at
$t= - \alpha_0$. It coincides with the usual Regge trajectory for
$t > - \alpha_0$ and with the large-$k$ zero-slope trajectory at
$t < - \alpha_0$.

It is intriguing to compare this behavior with the one emerging in the
holographic approach in Ref.\cite{Brower:2006ea} (for another holographically based approach to meson scattering amplitudes see also
\cite{Armoni:2016llq} and references therein). In the holographic
approach of \cite{Brower:2006ea}  there is only one string, however its slope depends
(continuously) on the radial coordinate making the slope varying
between 1 for small radial position (IR) to zero for large radial
position (UV). The effective outcome structure for the dominant
trajectory is again with the usual leading (IR) trajectory at large
positive $t$ and a "kink" at $t= - \alpha_0$ connecting to the constant
slope (UV) trajectory in the $t<0$ region. Figure 1 in
\cite{Brower:2006ea} clearly resembles our Fig.(1) and the mathematics
which leads to the power (up to logs) law behavior for fixed angle
scattering is quite similar. Of course in the holographic approach the
slope varies continuously with the radial position
and the computation of the amplitude
involves an integration over the radial coordinate while we
have just a discrete sum. Nevertheless the mathematical similarity leads in
both cases to the softer power (up to logs) behavior in the AF
region. If indeed the large-$N$ QCD will be proven to produce an
amplitude as in our model, it would be tempting to try and
understand/interpret the sum over $k$ as an indication for the
existence of an holographic dimension. It may very well happen that at
a certain level an interaction between the various strings
corresponding to the different terms in our sum will have to be put in
(without changing the qualitative behavior of the effective
$\alpha(t)$ in our model) thus turning the discrete index $k$ into a
continuous one corresponding to the holographic dimension.

To the extent that we take our model amplitude to be a first
approximation to the large-$N$ QCD amplitude we would like to
understand the origin of the sum over $k$ of string amplitudes with
slopes $1/k$. Large -$N$ QCD is believed to be some sort of tree-level
string theory. Moreover, lattice simulations \cite{Athenodorou:2010cs,
  Brandt:2016xsp} and the expansion of the effective theory of long
strings \cite{Aharony:2010db, Aharony:2013ipa} indicate that
this string theory should be rather close to the Nambu-Goto
string. The first term in our sum corresponds to such a string.  What
about the terms with $k >1$? One possible conjecture is that they
correspond to folded strings \cite{Ganor:1994bq} with the folding
occurring precisely at the end-points of the open string (where the
quark and anti-quark reside). Clearly, in order to account for the
folded string to go from the quark to the antiquark, the number of
folds must be even. The case of $2m$ folds corresponds to having a
tension of $k=2m+1$ (or slope of $1/(2m+1)$ in $\alpha'€™$
units) Hence, in order to account for such string configurations, the
sum in (\ref{GLS}) will be over odd integer $k$.

We have checked numerically that this still gives a unitary amplitude
and it still satisfies all the demands that we have imposed in section
2.  See Fig.s~(\ref{fig:PosOdd},\ref{fig:PosOddFlat}) where the allowed
region has been computed with the technique of
Sec.~\ref{sec:algebraic-algo}; notice that the lower bound
$\alpha_0\ge 1/2$ is due mainly to the $M=1$ data (as it is obvious
from Fig.\ref{fig:PosOdd})) for $\beta\ge \beta_0 \approx 1.198...  $,
while for smaller values of $\beta$ there are more stringent bounds
coming from higher values of $M$. For other values of $\gamma$ other
than $\beta+1$ we checked that the picture is essentially the same,
only the threshold $\beta_0$ is $\gamma$-dependent.

There is a-priori no reason we could think of why the folds should
occur at exactly the endpoints. Actually they could occur anywhere. We
wonder whether taking also such configurations into account will,
effectively, make the tension continuous and therefore the slope to
change continuously between one and zero. Having such a continuous
slope may make the formal resemblance to the holographic approach of
Ref.\cite{Brower:2006ea} even closer. 
\section*{Acknowledgements}
 
We would like to thank O. Aharony, N. Arkani-Hamed, A. Armoni, O. Ben-Ami,
M. Bochicchio, P. Di Vecchia, Z. Komargodski, A. Sever and M. Strikman
for interesting discussions. One of us (E.O.) would like to thank
Prof. N. Hale for useful correspondence about {\tt leg2cheb} and
P. Holoborodko for useful suggestions on our {\tt Advanpix/Matlab} code.  The
work of S.Y. is supported in part by the Israel Science Foundation
(ISF) Center of Excellence (grant 1989/14) ; by the US-Israel
bi-national science foundation (BSF) grant 2012383 and by the
German-Israel bi-national science foundation (GIF) grant
I-244-303.7-2013.

\newpage
\appendix
\section{Appendix}\label{sec:numerical-results}
We present here the strategy we used to expand the function given in
Eq.~\eqref{ResAx} into a series of Legendre polynomials; to be
specific we assume the coefficients in Eq.~\eqref{ResAx} to be
$c_k = k^{-\beta-1}\,(1+\ln(k))^{-\gamma}$ and typically, unless
differently specified, $\gamma=\beta+1$, as anticipated in
Sec.~\ref{sec:fixed-angle-limit}. The problem of Legendre expansion
received much attention in the last decades, and an account can be
found in Ref.s~\cite{AlpertRokhlin,HaleTownsend}. Fast techniques have
also been implemented in popular packages like Chebfun\footnote{{\tt
    www.chebfun.org}} working under Matlab\footnote{{\sl \copyright
    $1984$--$2017$ The MathWorks, Inc.}}. We had to recourse to a
different strategy because we need essentially unlimited arithmetic
accuracy to go beyond $M\approx O(100)$, and this is not possible with
Chebfun. The algorithms we are going to present\footnote{{\tt
    Mathematica} and {\tt Matlab} code employed here can be obtained
  by simply sending an email to the third author {\smallish{\it
      enrico.onofri}\at{\it unipr.it}}.} can be implemented either
in Mathematica\footnote{{\sl \copyright $2016$ Wolfram Research, Inc.,
    Champaign, IL (USA)}}, where one works with rationals and the
results are exact, or adopting a multiple precision software like {\tt
  advanpix}\footnote{{\tt www.advanpix.com}}.  We note here that an
entirely different approach could be adopted making recourse to a
classical result of Schoenberg \cite{Sch42}: a function
$f(z), (\minus 1<z<1) $ is ``of positive type on $S^2$'', if for any
set of unit vectors $\{v_j, j=1...n\}$ the matrix $f(v_i\cdot v_j)$ is
positive definite. Such a function of positive type has an expansion
in Legendre polynomials with non-negative coefficients (and
converse). We are exploring this property as an alternative to the
direct computation of the Legendre expansion by implementing a kind of
MonteCarlo on the space of $n$-tuples of unit vectors. Results will be
reported elsewhere.

We used both approaches, ``algebraic'' and ``spectral'', to derive the
unitarity region in $(\beta,\alpha_0)$. The first method has the
additional advantage that the calculations can be performed leaving
$\alpha_0$ and $\beta$ symbolic.  Since the calculation can be done in
unlimited precision we consider the numerical evidence as founded on
solid ground, and it is nice to have a perfect match with the analytic
saddle point technique.  Let us note that the spectral method can be
applied also to a non-polynomial $f(z)$ and this allowed to compare
the numerical result with that obtained by a saddle point technique in
Sec.\ref{sec:analytic-results} using Eq.\eqref{ImRegge}.

\subsection{The algebraic algorithm}\label{sec:algebraic-algo}
We have to expand a polynomial already given in factorized form, hence we can simply apply the
recursion relation of Legendre polynomials
\begin{equation}
\label{recursive}
  z\,P_n(z) = \frac{n+1}{2n+1}\,P_{n+1}(z) + \frac{n}{2n+1}\,P_{n-1}
\end{equation}
step by step. The problem is reduced to a simple algebraic
calculation, very easily implemented in a symbolic language like {\tt
  Mathematica}. The code starts with the vector representing $P_0(z)$,
$\phi_0=(1,0,0,\ldots,0,0)$ of length $M\plus2$ and one can apply the
various factors of the polynomial of Eq.\eqref{ResAx} by substituting
the variable $z$ with the matrix\footnote{A key ingredient for
  designing an efficient algorithm is the use of {\sl sparse matrix\/} techniques, otherwise the computation would grow as $O(M^3)$.}
  \begin{equation}
    Z=
    \begin{pmatrix}
      0&\frac13&0&0&0&0&\ldots\\
      1&0&\frac25&0&0&0&\ldots\\
      0&\frac23&0&\frac37&0&0&\ldots\\
      0&0&\frac35&0&\frac49&0&\ldots\\
      \vdots&&&\ddots&&\ddots&
    \end{pmatrix}
  \end{equation}
  as obtained by the recursion relation (\ref{recursive}) expressed in
  terms of the vector's components.  The final result is obtained by
  summing over all divisors of $M\plus1$ (or over {\sl odd\/} divisors for
  the case discussed in the conclusion section); we get the critical
  value for $a$ by a customary bisection method. This method is
  essentially exact, since step by step all components are rational
  numbers which can be treated with no truncation errors and the
  bisection method can be pushed to any desired accuracy.  The
  algebraic approach allowed us to explore the positivity region for
  values of $M$ well above $10^3$.  The only logical limitation is due
  to the fact that we can compute the coefficients only up to some
  finite value of $M$: how can we exclude that for some bigger value
  negative coefficients could possibly show up for the same values of
  the other parameters? We can't, but the overall picture and the
  asymptotic (analytic) formulae give us the confidence that this is
  not going to happen.

  \subsection{The spectral method}\label{sec:spectral-method}
  Another approach, allowing to get some additional insight into the
  problem, can be applied to {\sl any\/} function, not just to
  polynomials. If we restrict the expansion to a maximum order
  $n=M\plus1$ than we are working with a finite dimensional truncation
  of the matrix $Z$. Its spectral properties are well-known thanks to
  a result of Golub and Welsch about Gaussian quadrature
  \cite{GW1969}: its eigenvalues are given by the zeroes $z_j$ of
  $P_{M\plus2}(z)$ and its $j\mbox{-}$th eigenvector is given by
 $${\mathcal
   P}^{(j)}=\left\{\sqrt{w_j}\,P_n\left(z_j\right)\,\mid\,{\smallish
     n=0\ldots M\plus1}\right\}$$ where $\{w_j\}$ are the Gaussian weights
 for the $(M\plus2)\minus$dimensional quadrature rule. The zeros are all
 distinct and belong to the interval $(-1,1)$; the leading eigenvalue
 (i.e. the one closest to 1, by convention assigned to $j=1$)
 corresponds to an eigenvector with all positive components. This is
 clear from the properties of zeroes of orthogonal polynomials, but it
 is also an immediate consequence of Perron-Frobenius theorem about
 positive matrices (and our matrix is actually a {\sl stochastic\/}
 matrix). The expansion in Legendre polynomials is thus converted into
 an expansion in eigenvectors of the matrix $Z^{(M\plus2)}$, and for any
 function $f(z)$ we simply get (dropping for simplicity the index
 $M\plus2$)
$$
f(Z)_{n,m}\,=\, \sum_j\, f(z_j)\,{\mathcal P}^{(j)}_n \,{\mathcal P}^{(j)}_m
$$
The expansion coefficients we want to compute are obtained by applying
this matrix to the vector $\phi_0$ hence we simply get
$ C_n = f(Z)_{n,0}\equiv\sum_j\,w_j\,f(z_j)\,P_n(z_j) $, which is
nothing but Gauss' quadrature formula which is exact for polynomials
of degree $\le 2M\plus1$.  The expansion coefficients we have to
compute are obtained by identifying $f(z)$ with the r.h.s. of
Eq.\eqref{ResAx} or Eq.\eqref{saddlesum}, hence the spectral approach
is exact when applied to Eq.~\eqref{ResAx} while it is affected by an
error, which can in principle be estimated, in the other case.  For
$K=1$ the function is sharply peaked near $z=1$ so that the leading
eigenvector $j=1$ is dominant.  The situation is different for $k>1$
and the final result comes from a delicate balance of all these
contributions and it depends on $M$ and on the value of $\alpha_0$. So
we expect that for $\alpha_0$ sufficiently large the first eigenvector
with all positive components will be dominant, but only the explicit
calculation can identify the critical value of $\alpha_0$ above which
we get unitarity (see Figs.~\ref{fig:shaded} and \ref{fig:PosOddFlat}:
allowed values for $\alpha_0$ lie within the shaded region).


\end{document}